\documentclass[aps,prl,floatfix,twocolumn,groupedaddress]{revtex4}
\usepackage{graphicx}
\usepackage{epstopdf}
\usepackage{times}
\usepackage{amssymb,amsmath}
\usepackage{array}
\usepackage{multirow}
\usepackage{color}
\usepackage{setspace}
\usepackage{changes}
	\definechangesauthor[name={Per cusse}, color=orange]{per}

\usepackage{hyperref}
\hypersetup{hypertex=true,
    colorlinks=true,
    linkcolor=blue,
    anchorcolor=blue,
    citecolor=blue}
    
\usepackage{braket}
\usepackage{booktabs}
\usepackage{color}
\usepackage{threeparttable}
\usepackage{multirow}
\usepackage{mathtools}

\makeatletter

\begin{document}

\title[]{Heterogeneously integrated, superconducting silicon-photonic platform for measurement-device-independent quantum key distribution}

\author{Xiaodong Zheng$^{1,\dagger}$, Peiyu Zhang$^{1,\dagger}$, Renyou Ge$^{2,\dagger}$, Liangliang Lu$^{1,\dagger}$, Guanglong He$^{1,\dagger}$, Qi Chen$^{1,\dagger}$, Fangchao Qu$^{1}$, Labao Zhang$^{1,\ast}$, Xinlun Cai$^{2,\ast}$, Yanqing Lu$^{1}$, Shining Zhu$^{1}$, Peiheng Wu$^{1}$, Xiao-Song Ma$^{1,\ast}$}
\affiliation{
$^1$~National Laboratory of Solid-State Microstructures, Collaborative Innovation Center of Advanced Microstructures, School of Physics, Research Institute of Superconducting Electronics, School of Electronic Science and Engineering, Nanjing University, Nanjing 210093, China\\
$^2$~State Key Laboratory of Optoelectronic Materials and Technologies, School of Electronics and Information Technology, Sun Yat-sen University, Guangzhou 510000, China\\
$^{\dagger}$These authors contributed equally to this work\\
$^{\ast}$e-mails: lzhang@nju.edu.cn; caixlun5@mail.sysu.edu.cn; xiaosong.ma@nju.edu.cn}

\date{\today}

\begin{abstract}	
Integrated photonics provides a route both to miniaturize quantum key distribution (QKD) devices and to enhance their performance. A key element for achieving discrete-variable QKD is a single-photon detector. It is highly desirable to integrate detectors onto a photonic chip to enable the realization of practical and scalable quantum networks. We realize an integrated heterogeneous superconducting–silicon-photonic chip. Harnessing the unique high-speed feature of our optical waveguide-integrated superconducting detector, we perform the first optimal Bell-state measurement (BSM) of time-bin encoded qubits generated from two independent lasers. The optimal BSM enables an increased key rate of measurement-device-independent QKD, which is immune to all attacks against the detection system, and hence provides the basis for a QKD network with untrusted relays. Together with the time-multiplexed technique, we have enhanced the sifted key rate by almost one order of magnitude. With a 125~MHz clock rate, we obtain a secure key rate of 6.166~kbps over 24.0~dB loss, which is comparable to the state-of-the-art MDI-QKD experimental results with GHz clock rate. Combined with integrated QKD transmitters, a scalable, chip-based and cost-effective QKD network should
become realizable in the near future.
\end{abstract}

\maketitle
\onecolumngrid
\setstretch{1.667}
QKD employs the laws of quantum physics to provide information-theoretical security for key exchange~\cite{lo1999, gisin2002, scarani2009, xu2019, pir2019}. Despite the substantial progress in the past 35 years, practical implementations of QKD still deviate from ideal descriptions in security proofs, mainly due to potential side-channel attacks. For instance, a series of loopholes have been identified due to the imperfections of measurement devices~\cite{makarov2006, zhao2008, lydersen2010, elezov2019}. Inspired by the time-reversed entanglement-based QKD, measurement-device-independent QKD (MDI-QKD), which removes all detector side attacks, has been proposed~\cite{samuel2012,lo2012}. Instead of relying on the trusted nodes of traditional QKD protocols, MDI-QKD requires only a central node (Charlie) to perform a Bell-state measurement (BSM). The correlations between the two senders (Alice and Bob) can be obtained from the BSM results. Importantly, even if Charlie is not trusted, one can still guarantee the security of the MDI-QKD as long as Charlie can project his two photons onto Bell states. The outstanding features of MDI-QKD invite global experimental efforts, which are mainly based on bulk/fibre components~\cite{rubenok2013, liu2013,tang2014,tang2014experimental, wang2015, comandar2016, yin2016,wang2017, liu2019}. Despite the additional BSM by Charlie, the key rate~\cite{comandar2016} and the communication distance~\cite{yin2016} of MDI-QKD can be comparable with those of traditional QKD. Furthermore, the star-like topology of MDI-QKD quantum network is naturally suited for the metropolitan network with multiple users~\cite{frohlich2013,hughes2013,tang2016}. Recently the generalization of MDI protocol to multipartite schemes has been investigated~\cite{PhysRevLett.114.090501, zhu2015w, grasselli2019conference}.It has been shown that the performance of the multipartite schemes can be advantageous to iterative use of independent bipartite protocols\cite{grasselli2019conference}.

From the perspectives of hardware, recent developments involve particular integrated photonic devices for QKD, including on-chip encoders based on silicon modulators~\cite{ding2017,bunandar2018,ma2016,sibson2017,PhysRevX.10.031030}, on-chip transmitters including lasers, photodiodes, modulators based on indium phosphide (InP)~\cite{agnesi2019hong,Semenenko:20}, and decoders based on silicon oxynitride~\cite{sibson2017nc} and silicon dioxide~\cite{wang2019}, as well as integrated silicon-photonic chips for continuous-variable (CV) QKD~\cite{zhang2019,Tasker:2021rp}. The notion of MDI has also been extended to CV protocols~\cite{pirandola2015high} and can be applied for multipartite metropolitan network with considerable rate~\cite{ottaviani2019modular}. Most of the components used in QKD, including lasers, modulators and passive components (such as beam splitters and attenuators) are widely used in classical optical communication systems and are not specifically designed for QKD. In addition, single-photon detectors are indispensable for discrete-variable (DV) QKD system, because the senders' pulses have to have a mean photon number of less than 1 to guarantee communication security. So far, single-photon detector integrated chip platform has not been employed in a MDI-QKD system. In this work, we report the realization of a heterogeneous superconducting-silicon-photonic chip and its application for MDI-QKD.

\begin{figure*}[htbp]
\begin{center}
    \includegraphics[width=0.75\textwidth]{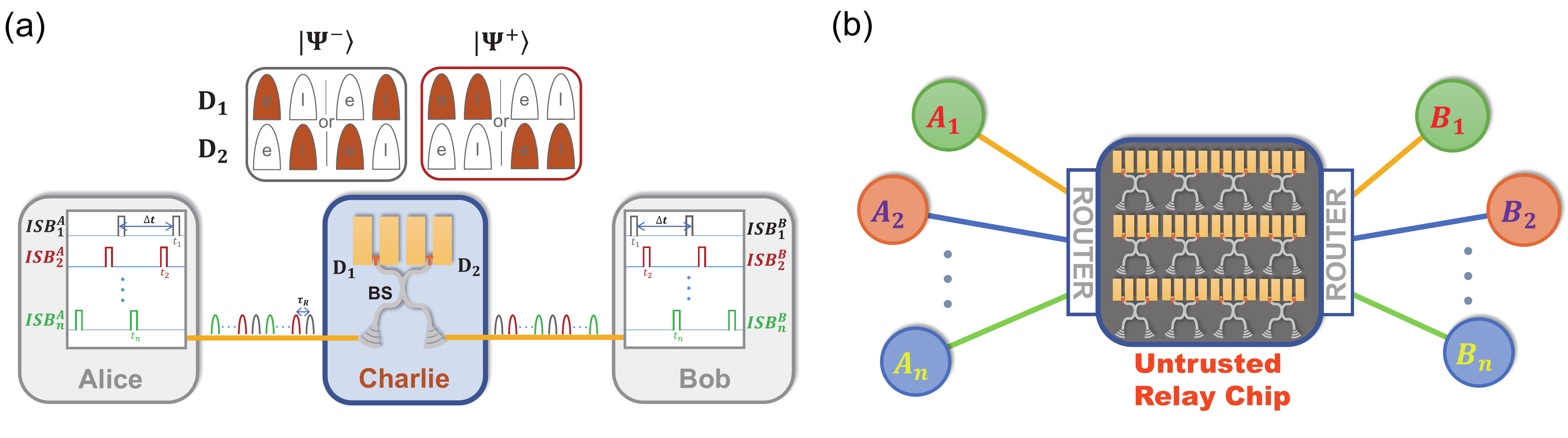}
    \caption{\label{Fig1} Schematic of a time-multiplexed MDI-QKD and a star-like MDI-QKD network: (a) Schematic of a time-multiplexed MDI-QKD with optimal Bell-state measurement (BSM). Alice and Bob send time-bin encoded qubits to Charlie for exchanging keys. By detecting the coincidence (red) between the early (e) and late (l) pulses with two detectors (D$_1$ and D$_2$), or with one detector (D$_1$ or D$_2$). Charlie projects two incoming photons onto $\ket{\Psi^{-}}$ or $\ket{\Psi^{+}}$ to facilitate the key exchanges between Alice and Bob. The full-recovery time of the single-photon detector sets the lower limit of the temporal separation ($\Delta t$) between e and l pulses for realizing optimal BSM.  We insert independent sets of bins (\textit{ISB}) between e and l for realizing time-multiplexing and hence increase the key rate by reducing the bin separation from $\Delta t$ to $\tau_{R}$. (b) A star-like MDI-QKD network with the untrusted relay server. A series of Alice (\textit{A}$_1$, \textit{A}$_2$, $\ldots$, \textit{A}$_n$) and Bob (\textit{B}$_1$, \textit{B}$_2$, $\ldots$, \textit{B}$_n$) prepare modulated weak coherent pulses and send to the routers. Two routers select a pair of Alice and Bob and send their pulses to an untrusted relay server controlled by Charlie.}
\end{center}
\end{figure*}

We use time-bin qubits to encode bit information, which are well suited for fibre-based quantum communication due to their immunity to random polarization rotations in fibres. The conceptual scheme of our experiment is shown in Fig. 1(a). Alice and Bob encode keys with time-bin qubits by using modulated weak coherent pulse sets. In Pauli Z-basis, the time-bins are encoded as the early, $\ket{e}$, and the late, $\ket{l}$, for bit values of 0 and 1, respectively. The temporal separation between $\ket{e}$ and $\ket{l}$ is $\Delta t$. In Pauli X-basis, the keys are encoded as the coherent superposition states between $\ket{e}$ and $\ket{l}$: $\ket{+}=(\ket{e}+\ket{l})/\sqrt{2}$ and $\ket{-}=(\ket{e}-\ket{l})/\sqrt{2}$, representing bit values of 0 and 1, respectively. The Z-basis code is used for key exchange, and the X-basis code is for error detection. These encoded time-bin qubits are then sent to Charlie, who performs the BSM on the incoming time-bin qubits by using a beam splitter (BS) and two single-photon detectors (D$_1$ and D$_2$)~\cite{samuel2012,lo2012}. By using linear optical elements, the success probability of BSM is bounded by 50\%~\cite{Calsamiglia:2001kq}. For projective measurements, optimal BSM corresponds to distinguish two out of four Bell states. Although time-bin qubits are well suited for fibre-based quantum communication, optimal BSM for time-bin qubits has yet to be realized. The bottleneck so far has been the lack of high-speed single-photon detectors~\cite{Houwelingen:2006nr,Semenenko:20,Samara2020}. The BSM scheme for time-bin qubits is shown in the inset of Fig. 1(a). The coincidence counts between two different detectors at different time bins, corresponding to coincidence counts between $\ket{e}_{D_1}$ (D$_1$ detects a photon at an early bin, red) and $\ket{l}_{D_2}$ (D$_2$ detects a photon at a late bin, red), or coincidence counts between $\ket{l}_{D_1}$ and $\ket{e}_{D_2}$. Such a coincidence detection projects two photons onto $\ket{\Psi^{-}}=(\ket{el}-\ket{le})/\sqrt{2}$, which is the common scenario realized in most of the time-bin BSM schemes~\cite{tang2014,Semenenko:20,Samara2020}. In order to achieve optimal BSM, we also need to detect $\ket{\Psi^{+}}=(\ket{el}+\ket{le})/\sqrt{2}$ by measuring the coincidence counts of one detector at different time bins, corresponding to the coincidence detection between $\ket{e}_{D_1}$ and $\ket{l}_{D_1}$, or $\ket{e}_{D_2}$ and $\ket{l}_{D_2}$. This particular BSM requires high-speed single-photon detection, capable to detect consecutive photons separated by $\Delta t$. The unique design of the waveguide-integrated SNSPD provides a short recovering time ($\textless$10 ns) for single-photon detection, enabling us to perform time-bin-encoded optimal BSM between two independent lasers for the first time. Note that if we only use one set of time-bin qubits, the system repetition rate will be limited to $1/(2\Delta t)$. In order to maximize the channel efficiency, we use time-multiplexed encoding to insert independent sets of bins (\textit{ISB}$^{\textit{A}}_2$, ..., \textit{ISB}$^{\textit{A}}_n$ and \textit{ISB}$^{\textit{B}}_2$, ..., \textit{ISB}$^{\textit{B}}_n$) between the $\ket{e}$ and $\ket{l}$ bins of \textit{ISB}$^{\textit{A}}_1$ and \textit{ISB}$^{\textit{B}}_1$. Therefore, the system repetition rate will be greatly increased to $1/(2\tau_{R})$, where $\tau_{R}$ is the time difference between $t_{1}$ and $t_{2}$. By harnessing the optimal BSM and time-multiplexing, the key rate generation is enhanced by an order of magnitude compared to the system without using these two techniques. Consequently, our key rate is comparable to the state-of-the-art MDI-QKD experimental results with GHz clock rate, as detailed later.

Our integrated heterogeneous superconducting-silicon-photonic platform provides a server architecture for realizing a multiple-user trust-node-free quantum network with a fully-connected bipartite-graph topology. As shown in Fig. 1(b), modulated weak coherent pulses are prepared by Alices (\textit{A}$_1$, \textit{A}$_2$, $\ldots$, \textit{A}$_n$) and Bobs (\textit{B}$_1$, \textit{B}$_2$, $\ldots$, \textit{B}$_n$), and are sent to the routers. Two routers select the pair of the communicating Alice and Bob, and send their pulses to an untrusted relay server controlled by Charlie. At Charlie's station, a chip with multiple low-dead-time~\cite{pernice2012}, low-timing-jitter~\cite{korzh2018} and high-efficiency detectors in conjunction with low-loss silicon photonics~\cite{ferrari2018} are used to realize the BSM. This configuration allows any user at Alice's side to communicate with any user at Bob's side and hence to realize a fully-connected bipartite quantum network. 

The schematic of our experimental setup is shown in Fig. 2(a). Alice (Bob) chops the CW laser operated at about 1536.47 nm into desired pulse sequences. The pulse is about 370 ps wide and separated by 12 ns at a rate of 41.7 MHz rate (~1/24~ns). Z-basis (X-basis) states are generated by chopping the laser into $\ket{e}$ or (and) $\ket{l}$ states with intensity modulators (IMs). The average photon numbers per pulse in the two bases are about the same. The resulting pulses are sent into a phase modulator (PM) with (without) $\pi$-phase shift for the generation of $\ket{-}$ ($\ket{+}$) states. The electrical signals applied to the modulators are generated by an arbitrary waveform generator (AWG, not shown in Fig. 2(a)). An additional 50:50 beam splitter (BS) combined with a power sensor (PS) is employed to monitor the long-term stability of laser power in each encoder. 

\begin{figure*}[htbp]
\begin{center}
    \includegraphics[width=0.75\textwidth]{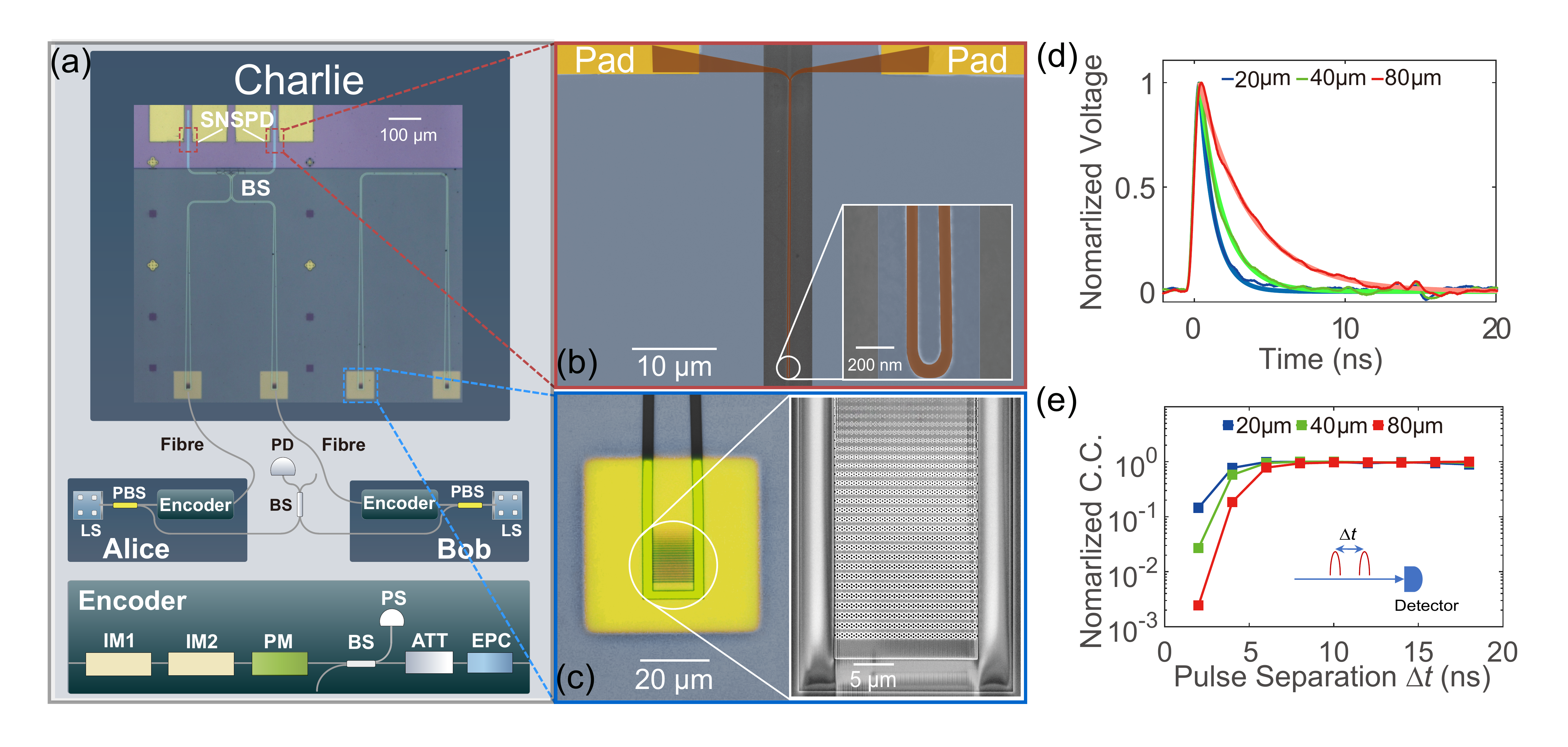}
    \caption{\label{Fig2} Experimental device and setup: (a) Schematic of the experiment setup. Alice (Bob) employs a continuous-wave laser as the light source (LS) and encodes the keys into optical pulses with an encoder module. In this module, one intensity modulator (IM1) chops out early ($\ket{e}$) and late ($\ket{l}$) temporal modes to generate time-bin qubits with 370 ps duration and separated by 12 ns with 41.7 MHz repetition rate. IM2 implements intensity modulation for the decoy-state protocol. A phase modulator (PM) applies a $\pi$-phase to the late temporal modes for $\ket{-}$ and 0-phase for $\ket{+}$ in X-basis. This PM also implements the phase randomization required for MDI-QKD. A variable attenuator (ATT) prepares weak coherent pulses and simulates the propagation loss in fibres. An electrical polarization controller (EPC) adjusts the polarization of input pulses. The pulses travel through fibres and are coupled into the integrated chip of the relay server (Charlie) for BSM. On the chip, we use a multi-mode interferometer acting as a 50:50 beam splitter (BS) and two SNSPDs. (b) False-colour scanning electron micrograph (SEM) of the SNSPD. A 80 nm-wide, 80 $\mu$m-long U-shaped NbN nanowire is integrated on a 500 nm-wide silicon optical waveguide and connected with two gold pads for electrical readout. The inset shows the zoomed part of the nanowire. (c) Optical and SEM graphs of the high-efficiency photonic-crystal grating coupler with a back-reflected mirror. (d) The averaged amplified response pulses of the 80-nm-wide SNSPD with different length. The 1/e-decay time of different SNSPD is obtained by fitting: 20 $\mu$m -- 0.96 ns; 40 $\mu$m -- 1.56 ns; 80 $\mu$m -- 3.39 ns. (e) Normalized coincidence counts of one detector consecutively detecting both early and late time bins as a function of time separation $\Delta t$ between them. Abbreviations: PBS, polarization beam splitter; PD, photodiode; PS, power sensor; EPC, electrical polarization controller.}
\end{center}
\end{figure*}

One of the most important requirements of MDI-QKD is to obtain high-quality two-photon Hong--Ou--Mandel (HOM) interference on the integrated relay server. To achieve that, it is necessary for Alice and Bob to generate indistinguishable weak-coherent pulses. The interfering pulses have to be indistinguishable in all degrees of freedom (DOF), including spectrum, time, and polarization. For the spectrum DOF, Alice's and Bob's unmodulated pulses pass through polarization beam splitters (PBSs), with one of the outputs connected with a 50:50 beam splitter (BS) for frequency beating. From the beating signal, we feedback onto one of the lasers and regulate the frequency difference of these two lasers to be within 10 MHz (See Supplementary Material for details). For the polarization DOF, two electrical polarization controllers (EPC) are used to optimize the polarization of both pulses before they are coupled into Charlie's chip. For the temporal DOF, we adjust the relative electrical delay between Alice's and Bob's IMs to ensure that their pulses arrive at the chip simultaneously. Attenuators are used to adjust the average photon number per pulse and simulate the loss of the communication channels. 

These pulses are then sent to Charlie's relay server chip which is mounted on a nano-positioner in a closed-cycle cryostat with a base temperature of 2.1 K. We show the U-shape waveguide-integrated superconducting nanowire single-photon detector (SNSPD) in Fig. 2(b), in which the superconducting nanowire (80 nm-wide, 80 $\mu$m-long) is highlighted in red and the silicon optical waveguide (500 nm-wide) is shown in blue. Fig. 2(c) shows the scanning electron microscope image of the photonic-crystal grating coupler~\cite{ding2014, luo2018}, which couples light from the fibre array to the chip. We obtain coupling loss from the reference device, which is at the right side of the main device~\cite{Gaggero:19}. The grating coupler with back-reflected mirror offers a coupling loss of $\sim$2.24~dB at a wavelength of 1536 nm. The main device has two identical grating couplers, coupling Alice's and Bob's pulses from fibre to chip. Silicon optical waveguides guide the pulses to a multi-mode interference (MMI) coupler, which acts as a 50:50 beam splitter. At the output of the MMI, two waveguide-integrated SNSPDs work simultaneously for detecting photons. Both SNSPDs are biased with constant voltage sources and connected with electronic readout circuitries. In Fig. 2(d), we show the electrical signals of waveguide-integrated SNSPDs with different nanowire length. The decay time of SNSPD is directly proportional to the kinetic inductance of the nanowire. Shorter detectors exhibit lower kinetic inductance and therefore have shorter decay times, resulting in faster detector recovery~\cite{Kerman:2006cv}. However, for traditional normal-incidence SNSPDs, the shorter nanowire length leads to lower detection efficiency, because it is necessary to fabricate large-area meander nanowire to match the optical modes from fibres to obtain high detection efficiency.Therefore, it is hard to simultaneously obtain low dead time and high detection efficiency with the traditional design. In our work, we use the evanescent coupling between optical waveguide and superconducting nanowire to circumvent this trade-off. Therefore, we are able to obtain low dead time as well as high on-chip detection efficiency. To further quantitatively characterize the efficiency of our SNSPDs for projecting two photons onto $\ket{\Psi^{+}}$, we measure the normalized coincidence counts of one detector consecutively detecting both early and late time bins as a function of time separation $\Delta t$ between them. The experimental results are shown in Fig. 2(e). The detection probability is significantly decreased when the time separation is smaller than the dead time, and is fully recovered for separation larger than 12 ns. Based on these results, the dead time of the SNSPD we use in our QKD system is about 3.4 ns for the 1/e-decay time, and we set the time separation between $\ket{e}$ and $\ket{l}$ to be 12 ns. This short time separation not only allows high-speed detection, but also greatly simplifies frequency stabilization of the light source. For a traditional normal-incidence SNSPD which limits 75 ns time-bin separation~\cite{tittel2014}, a 185 kHz frequency difference between two lasers can result in $5^{o}$ phase error, which is technologically challenging and not practical. By contrast, for our waveguide-integrated SNSPD, the frequency-stabilization requirement is only 1.2 MHz for achieving the same phase error, which is significantly more feasible in practice. 

\begin{figure*}[htbp]
\begin{center}
    \includegraphics[width=0.8\textwidth]{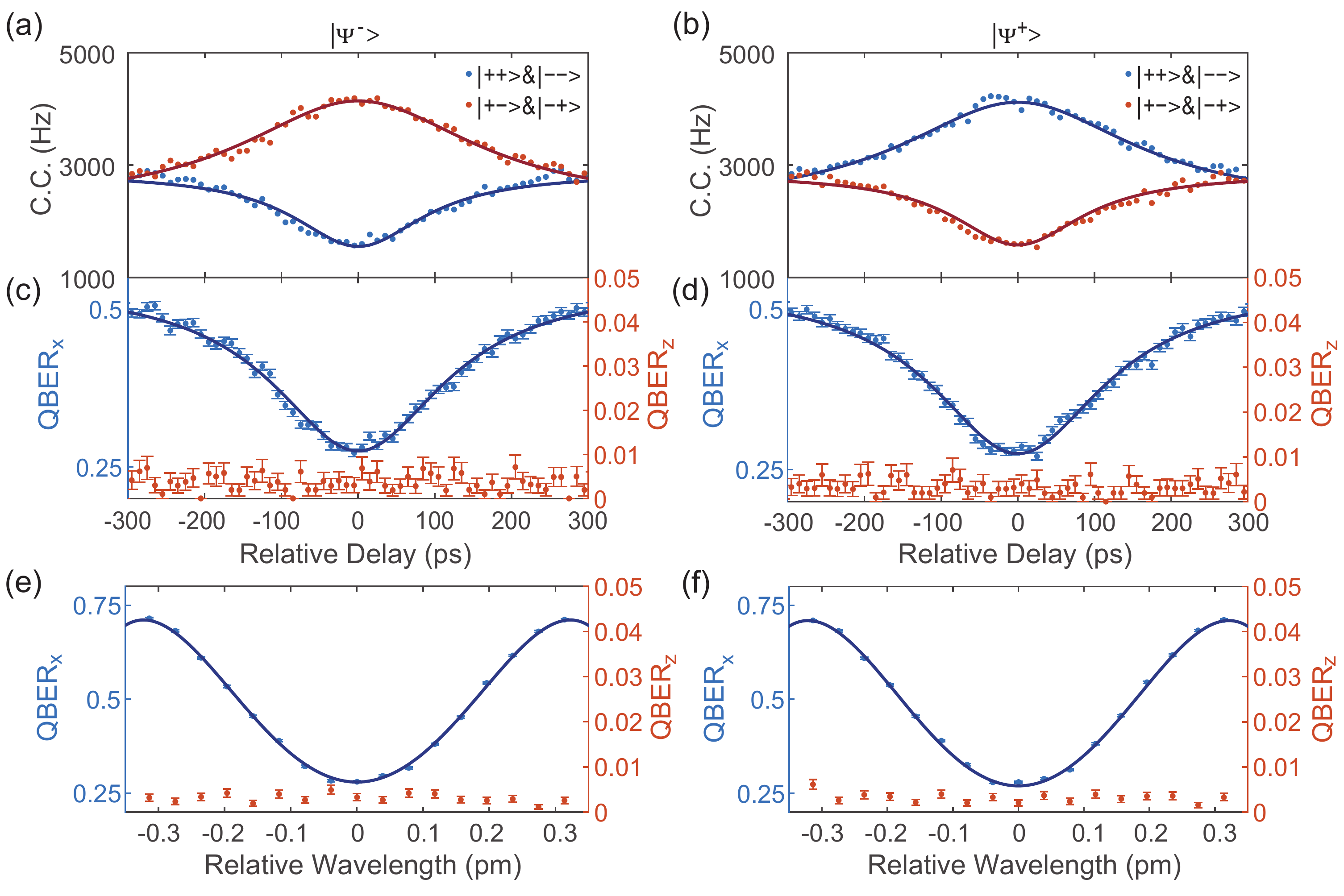}
    \caption{\label{Fig3} Experimental results of optimal BSM and quantum bit error rate (QBER). (a) BSM results of $|\Psi^{-}\rangle$. When Alice and Bob send the same states ($\ket{++}$/$\ket{--}$, blue dots), or different states ($\ket{+-}$/$\ket{-+}$, red dots), we obtain destructive and constructive interference in coincidence counts as functions of relative temporal delay, respectively. (b) BSM results of $|\Psi^{+}\rangle$. Note that the correlations between Alice and Bob are inverted comparing to $|\Psi^{-}\rangle$. (c) and (d) show the QBER in X-basis (blue) and Z-basis (red) for $|\Psi^{-}\rangle$ and $|\Psi^{+}\rangle$, respectively. (e) and (f) show the measured QBER in X-basis and Z-basis as a function of the wavelength detuning between two lasers for two different Bell states.}
\end{center}
\end{figure*}

In Fig. 3(a) and (b), we show the two-photon coincidence counts with optimal BSM as a function of relative electronic delays between Alice's and Bob's pulse sequence, in which Charlie projects the two photons sent by Alice and Bob onto $\ket{\Psi^{-}}$ and $\ket{\Psi^{+}}$, respectively. The dependence of the coincidence counts on the delay is a result of BSM, showing the coherent two-photon superposition. Due to the symmetry of $\ket{\Psi^{-}}$ and $\ket{\Psi^{+}}$, when Alice and Bob send the same states in X-basis, $\ket{++}$ or $\ket{--}$, we obtain the destructive/constructive interference patterns for the BSM results of $\ket{\Psi^{-}}$/$\ket{\Psi^{+}}$, as shown by the blue dots in Fig. 3(a) and (b). When Alice and Bob send the orthogonal states in X-basis, $\ket{+-}$ or $\ket{-+}$, we obtain the inverse results, as shown by the red dots in Fig. 3(a) and (b). (The logic of coincidence detection for $\ket{\Psi^{-}}$ and $\ket{\Psi^{+}}$ is shown in the inset of Fig. 1(a).)

We obtain secure keys from the Z-basis measurements and verify the reliability of the QKD system in X-basis~\cite{yin2016}. To quantify the performance of the system, we analyze the quantum bit error rate (QBER). For instance, Alice and Bob exchange their keys conditionally on Charlie obtaining $|\Psi^{-}\rangle$/$|\Psi^{+}\rangle$ from his BSM, when Alice and Bob send the same/orthogonal states. For X-basis, the probability of Charlie obtaining a coincidence at two subsequent time bins with time separation $\Delta t$ is $P_X^{-}(t,t+\Delta t)$/$P_X^{+}(t,t+\Delta t)$. We then obtain the QBER in X-basis ($QBER_X^{|\Psi^{-}\rangle}$/$QBER_X^{|\Psi^{+}\rangle}$) based on~\cite{jin2013}:
\begin{equation} \label{QBER_X}
QBER_X^{|\Psi^{-}\rangle}=\frac{P_X^{-}(t,t+\Delta t)}{P_X^{+}(t,t+\Delta t)+P_X^{-}(t,t+\Delta t)}.
\end{equation}
\begin{equation} \label{QBER_X}
QBER_X^{|\Psi^{+}\rangle}=\frac{P_X^{+}(t,t+\Delta t)}{P_X^{+}(t,t+\Delta t)+P_X^{-}(t,t+\Delta t)}.
\end{equation}
\begin{equation} \label{QBER_Z}
QBER_Z^{|\Psi^{\pm}\rangle}=\frac{P_Z^{-}(t,t+\Delta t)}{P_Z^{+}(t,t+\Delta t)+P_Z^{-}(t,t+\Delta t)}.
\end{equation}
In addition, the phase difference of two subsequent time bins induced by frequency difference is
\begin{equation} \label{GrindEQ2}
\theta=2\pi\Delta \omega \Delta t=2\pi(\omega_a-\omega_b)\Delta t=2\pi c\Delta t\frac{|\lambda_a-\lambda_b|}{\lambda_a\lambda_b}.
\end{equation}
where $c$ is the speed of light, $\omega_a$ ($\omega_b$) and $\lambda_a$ ($\lambda_b$) are the laser's frequency and wavelength of Alice (Bob), respectively.

$P_X^{-}(t,t+\Delta t)$/$P_X^{+}(t,t+\Delta t)$ can be written as:
\begin{equation} \label{GrindEQ3}
P_X^{\pm}(t,t+\Delta t)=1\pm V\exp[-\tau^2(c\frac{|\lambda_a-\lambda_b|}{\lambda_a\lambda_b})^2]\cos\theta.
\end{equation}
where $V$ is the visibility, $\tau$ is the coincidence window. 

As for Z-basis, $QBER_Z$ always have the same formula for $|\Psi^{-}\rangle$/$|\Psi^{+}\rangle$. In Fig. 3(c) and (d), we show the measured $QBER_X^{|\Psi^{-}\rangle}$ and $QBER_X^{|\Psi^{+}\rangle}$ (blue) as functions of time delays between Alice and Bob, which show the minimum close to 0.25 at the zero time delay. For Z-basis, the measured $QBER_Z^{|\Psi^{\pm}\rangle}$ (red) are close to zero, showing the high quality of our system. In Fig. 3(e) and (f), we vary the relative central wavelength between Alice’s and Bob’s lasers. And we show the results for $QBER_X^{|\Psi^{-}\rangle}$ and $QBER_X^{|\Psi^{+}\rangle}$ as functions of the relative wavelength, respectively. The experimental data (blue dots) agree well with the theoretical prediction (blue curves).

\begin{figure*}[htbp]
\begin{center}
    \includegraphics[width=0.5\textwidth]{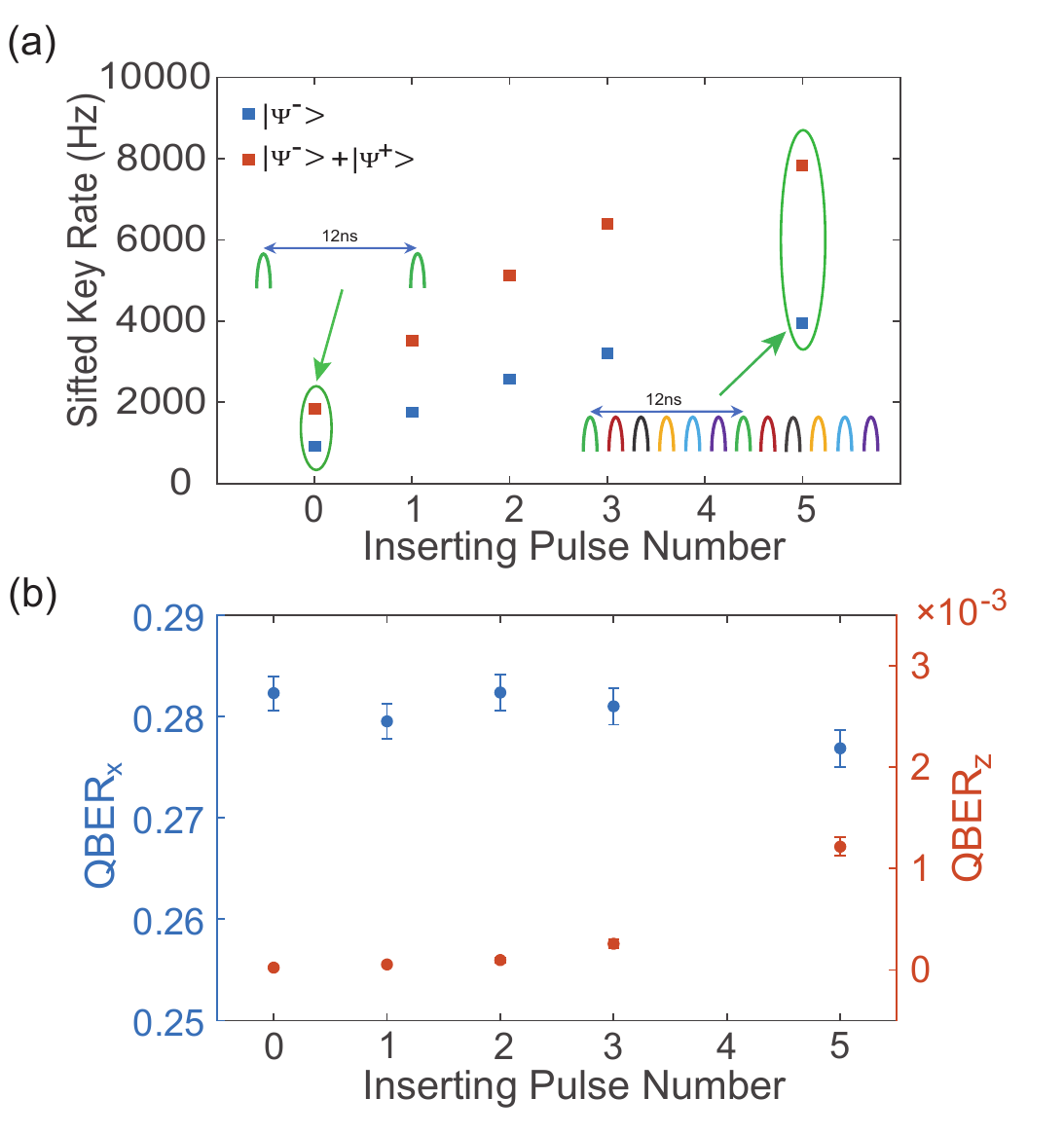}
   \caption{\label{Fig4} Enhanced key rate by time-multiplexing. (a) The sifted key rate as a function of the inserted pulse number between the full-recovery time of SNSPD (12 ns). Red squares are the results of optimal BSM and blue squares are the result of $|\Psi^{-}\rangle$-only measurement. To compare fairly, in all the results presented here, Alice and Bob send the weak coherent pulses with the average photon number of 0.66 per pulse, and the total loss is 35.0~dB (including chip insertion loss $\sim$~4.5~dB). (b) $QBER_X$ and $QBER_Z$ versus inserting pulse number, indicating that time-multiplexing has little influence on error rate.}
\end{center}
\end{figure*}

Although the full-recovery time of the detector determines the time-bin separation to be 12 ns, we can harness the time-multiplexed technique by inserting more pairs of time-bin pulses to enhance the key rate. This is particularly useful in high-loss communication applications. As shown in the insets of Fig. 4(a), we insert up to five bins between 12 ns with equal temporal separation of 2 ns. By combining this time-multiplexed technique and optimal BSM, we enhance the sifted key rate by almost an order of magnitude. At the same time, these two techniques have little impact on $QBER_X$ and $QBER_Z$, as shown in Fig. 4(b).

\begin{figure*}[htbp]
\begin{center}
    \includegraphics[width=0.5\textwidth]{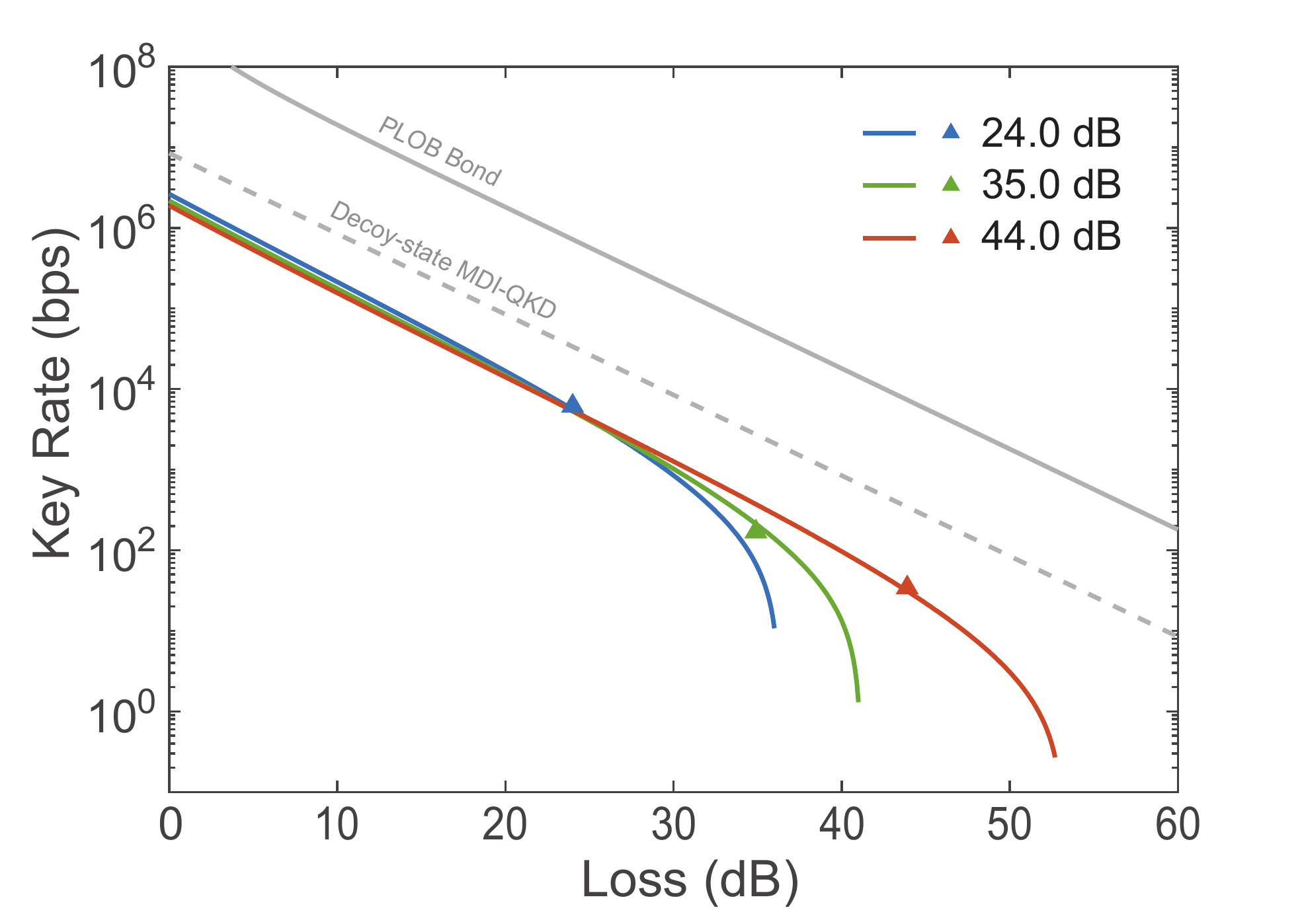}
    \caption{\label{Fig5} The key rate at different loss including chip insertion loss. The solid lines show theoretical simulations and the triangle symbols show experimental results with a  loss of 24.0~dB, 35.0~dB and 44.0~dB, respectively. For different loss, the paremeters (the intensities, $s$, $\mu$, $\nu$, and the probability of intensities, $P_s$, $P_{\mu}$, $P_{\nu}$) are different. See Supplementary Material for detailed paremeters of theoretical simulations. The gray solid line - PLOB bond~\cite{pirandola2017fundamental} and the gray dot line - Decoy-state MDI-QKD are numerical simulations. See Supplementary Material for details.}
\end{center}
\end{figure*}

We demonstrate a complete MDI-QKD system including decoy states and phase randomization for guaranteeing the security~\cite{liu2013, tang2014, tang2014experimental, wang2015, comandar2016, yin2016, zhou2016, wang2017, liu2019, zhang2017} with our integrated heterogeneous superconducting-silicon-photonic platform. We use a four-intensity encoding protocol~\cite{zhou2016} with three intensities ($\mu$, $\nu$, $o$) in the X-basis for decoy-state analysis and one intensity ($s$) in the Z-basis for key generation. Finite-key effects are considered in the secure-key-rate analysis with a failure probability of 10$^{-10}$~\cite{curty2014}. For statistical fluctuations, we use the joint constrains where the same observables are combined and treated together~\cite{zhou2016} (See Supplementary Material for details). 

In this part of the experiment, we evenly insert two more pairs of time-bin qubits between 12~ns separation. Therefore, the effective clock rate of our system is tripled to 125~MHz (1/8~ns). The secure key rates for different losses are shown in Fig. 5. With 125 MHz clock rate, we obtain the key rate of 6.166~kbps at the loss of 24.0~dB. This loss includes chip insertion loss $\sim$~4.5~dB. The actual transmission loss is about 19.5~dB, which corresponds to 98 km standard fibre. To the best of our knowledge, this is the highest secure key rate obtained experimentally with $\sim$20~dB transmission loss in MDI-QKD, which is highly relevant in the context of a metropolitan quantum network without detector vulnerabilities. Furthermore, we obtain the secure key rates of 170~bps and 34~bps with the total losses of about 35.0~dB and 44.0~dB. We emphasize that our secure key rates with 125~MHz clocked system are very close to the best MDI-QKD experiments with GHz clock rate~\cite{PhysRevX.10.031030, woodward2021gigahertz}. In contrast with GHz clock rate MDI-QKD experiments, our system doesn't require complicated injection locking technique, which significantly reduces the complexity of the transmitter. See the table in Supplementary Material for detailed comparison.

In conclusion, we have demonstrated the first integrated relay server for MDI-QKD with a heterogeneous superconducting-silicon-photonic chip. The excellent optical and electronic performance of this chip not only facilitates the experimental high-visibility HOM interference and low QBER, but also allows us to perform optimal BSM for time-bin qubits for the first time. Our work shows that integrated quantum-photonic chips provides not only a route to miniaturization, but also significantly enhance the system performance than traditional platforms. Our chip-based relay server can also be employed in Twin-Field QKD (TF-QKD) \cite{lucamarini2018overcoming} which can overcome the rate-distance limit of QKD without quantum repeaters. TF-QKD is indispensable in long-distance intercity communication link. Moreover, the chip-based relay server with MDI-QKD protocol presented in this work could be an ideal solution for a scalable trust-node-free metropolitan quantum network. By using more advanced waveguide-integrated SNSPDs~\cite{ferrari2018}, one can further improve the integrated server with high detection efficiency, low timing jitter, and high repetition rate. Combined with photonic-chip transmitters~\cite{Semenenko:20, PhysRevX.10.031030}, a fully chip-based, scalable and high-key-rate MDI-QKD metropolitan quantum network should be realized in the near future.

\section*{Acknowledgements}
We thank R. Chen and A. Miller for helpful discussions. This research is supported by the National Key Research and Development Program of China (2017YFA0303704, 2019YFA0308700 and 2017YFA0304002), National Natural Science Foundation of China (Grants No. 11690032, No. 11321063 and 12033002), NSFC-BRICS (No. 61961146001), Leading-edge technology Program of Jiangsu Natural Science Foundation  (BK20192001), and the Fundamental Research Funds for the Central Universities.\noindent

\noindent
\bibliographystyle{naturemag}
\bibliography{MDI_ref}

\clearpage
\newpage

\section{Supplementary information}

\subsection{The heterogeneous superconducting-silicon-photonic chip}\label{sec:one}
In this section, we introduce the fabrication and characterization of the heterogeneous superconducting-silicon-photonic chip, including high-efficiency grating couplers, optical waveguides, multimode interference (MMI) couplers and waveguide-integrated superconducting nanowire single-photon detectors (SNSPDs).

The six fabrication steps of the high-efficiency grating couplers are shown in Fig. \ref{Fig1}.
\begin{figure*}[htbp]{}
\begin{center}
    \includegraphics[width=0.8\textwidth]{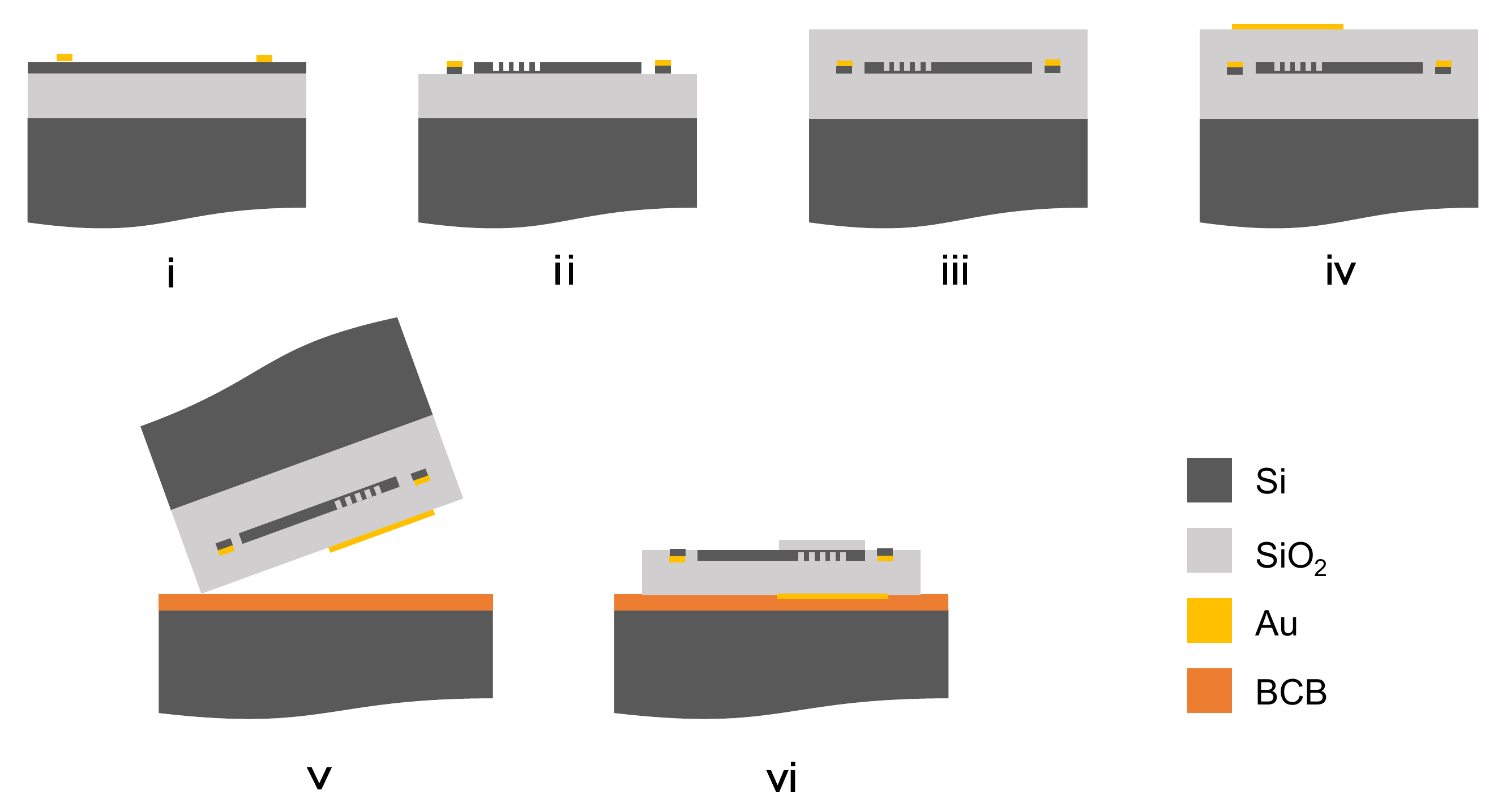}
    \caption{\label{Fig1} Six-step fabrication process of the high-efficiency grating couplers. (i) The device is fabricated on a commercial silicon-on-insulator (SOI) wafer with 3 $\mu$m thick buried oxide layer and 220 nm thick top silicon layer. We fabricate gold (Au) pattern as the alignment markers for the subsequent steps. (ii) Silicon-photonic structures (including waveguides, MMIs and grating couplers) are patterned using electron beam lithography and etched using the inductively coupled plasma etching process. (iii) A thickness of 3 $\mu$m $SiO_2$ cladding is then deposited via the plasma-enhanced chemical vapor deposition and planarized via chemical mechanical polishing to an optimized thickness. (iv) Gold reflectors are then defined by lithography, metal evaporation and lift-off processes on the $SiO_2$ cladding right above the grating couplers. (v) The chip is then flip-bonded to a silicon carrier wafer using benzocyclobuten (BCB). (vi) Finally, silicon substrate of the chip is removed by mechanical grinding and chemical etching, followed by dry etching the buried oxide layer.}
\end{center}
\end{figure*}

Fig. \ref{Fig2} shows the transmission spectrum of a reference device of $\sim$ 1500-$\mu$m-long waveguide with two grating couplers as the input and output, respectively. The total 
insertion loss of this reference device is about -4.48 dB at a wavelength of 1536 nm. If we neglect the relative small propagation loss of the on-chip waveguide, the insertion loss of each fiber-to-chip grating coupler is -2.24 dB at 1536 nm.

\begin{figure*}[htbp]
\begin{center}
    \includegraphics[width=0.8\textwidth]{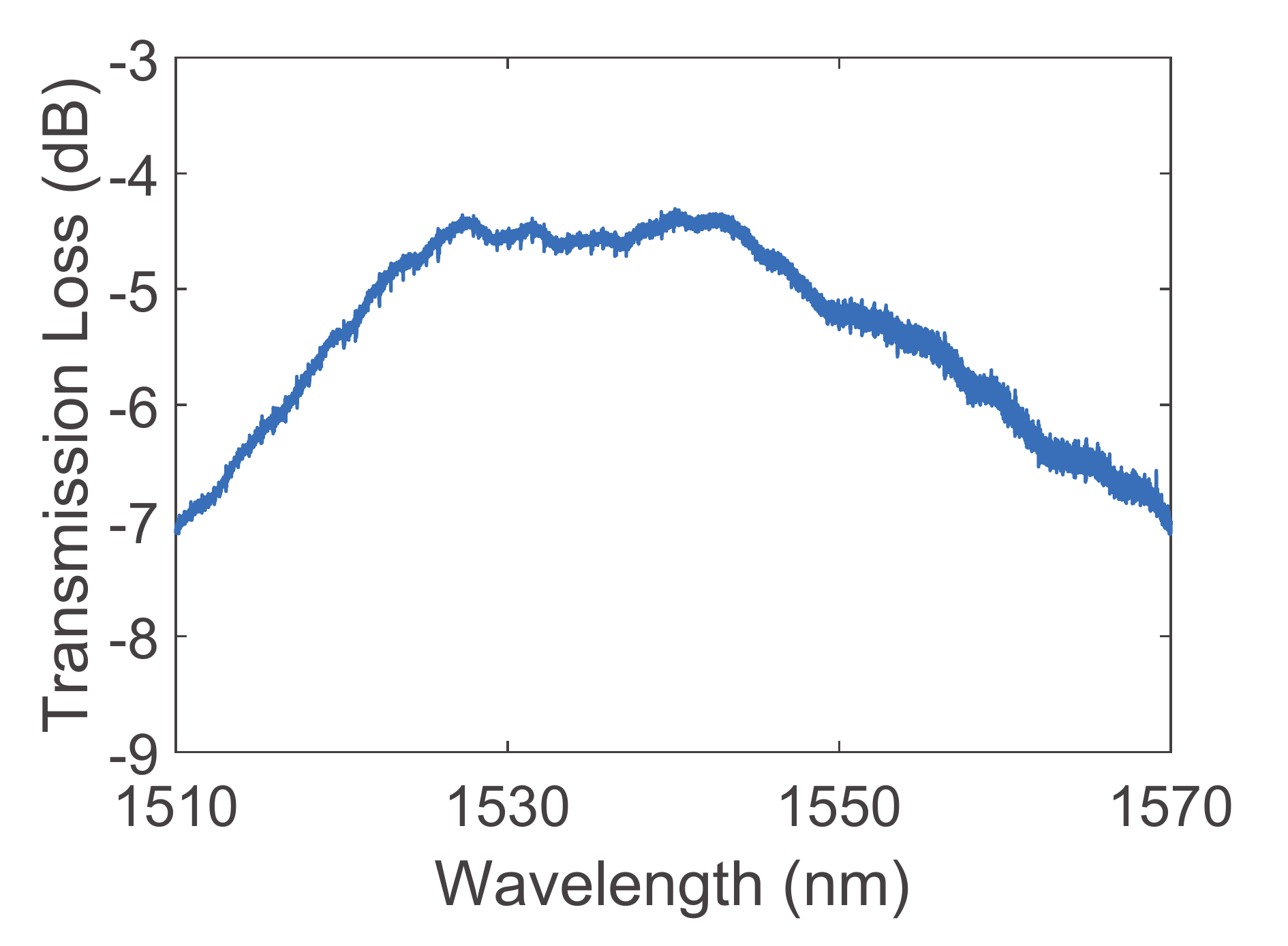}
    \caption{\label{Fig2} The transmission spectrum of the reference device measured in a cryogenic setup.}
\end{center}
\end{figure*}

\begin{figure*}[htbp]
\begin{center}
    \includegraphics[width=0.8\textwidth]{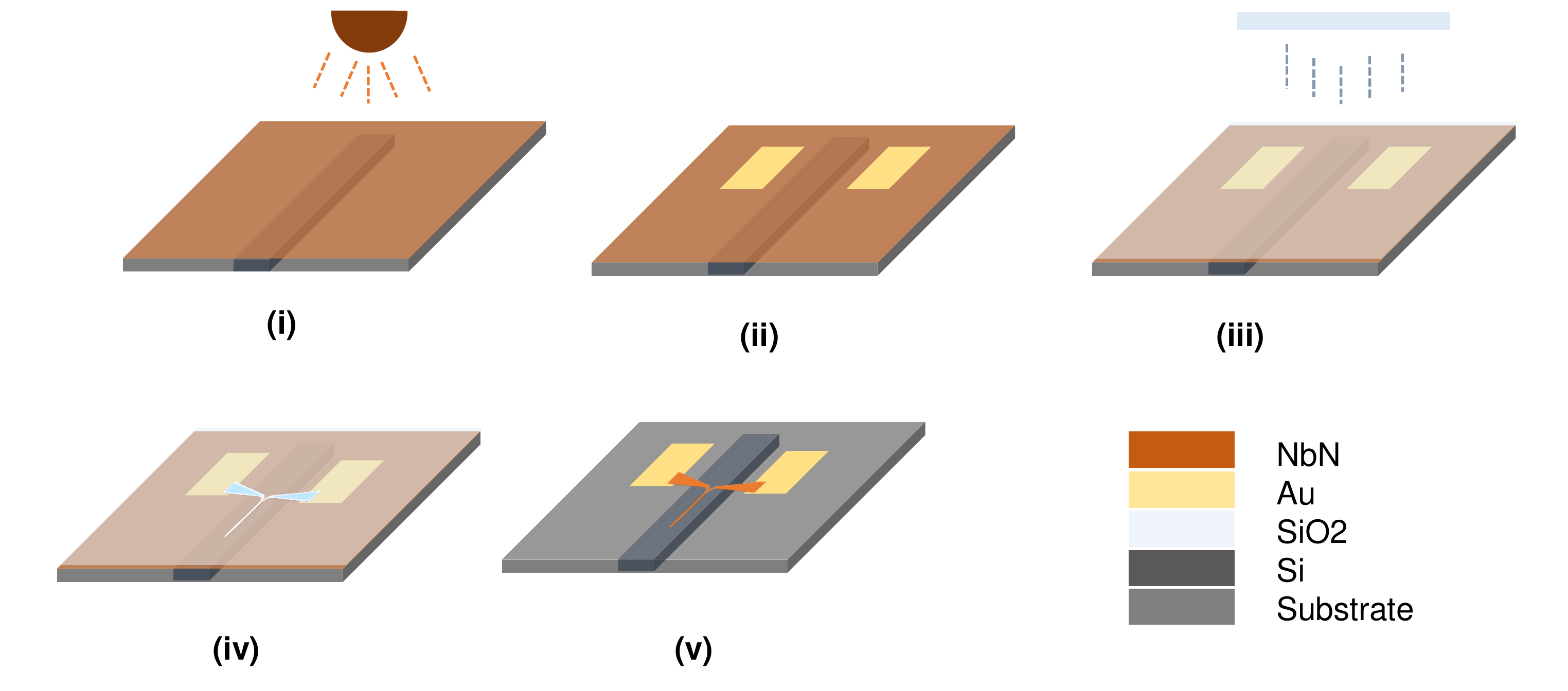}
    \caption{\label{Fig3} Fabrication of NbN based SNSPD. (i) A nominal 6 nm NbN film is deposited on the chip by DC reactive magnetron sputtering. Sheet resistance is R$_s$=341 $\Omega/\Box$, and critical temperature is about 6.1 K. (ii) A series of fabrication process includes electron beam lithography (EBL), Titanium (Ti) and Gold (Au) deposition and lift-off, are used to fabricate gold pads for electronic contact. (iii) A 10 nm $SiO_2$ layer is deposited on NbN film by plasma-enhanced chemical vapor deposition to improve the adhesion between 2\% hydrogen silsesquioxane (HSQ) and NbN. (iv) Then, the detector pattern is defined by EBL and developed using 2.38\% TMAH. (v) Lastly, $SiO_2$ and the unwanted NbN are etched away in a timed reactive ion etching step with $CF_4$ plasma.}
\end{center}
\end{figure*}

\begin{figure*}[htbp]
\begin{center}
    \includegraphics[width=0.8\textwidth]{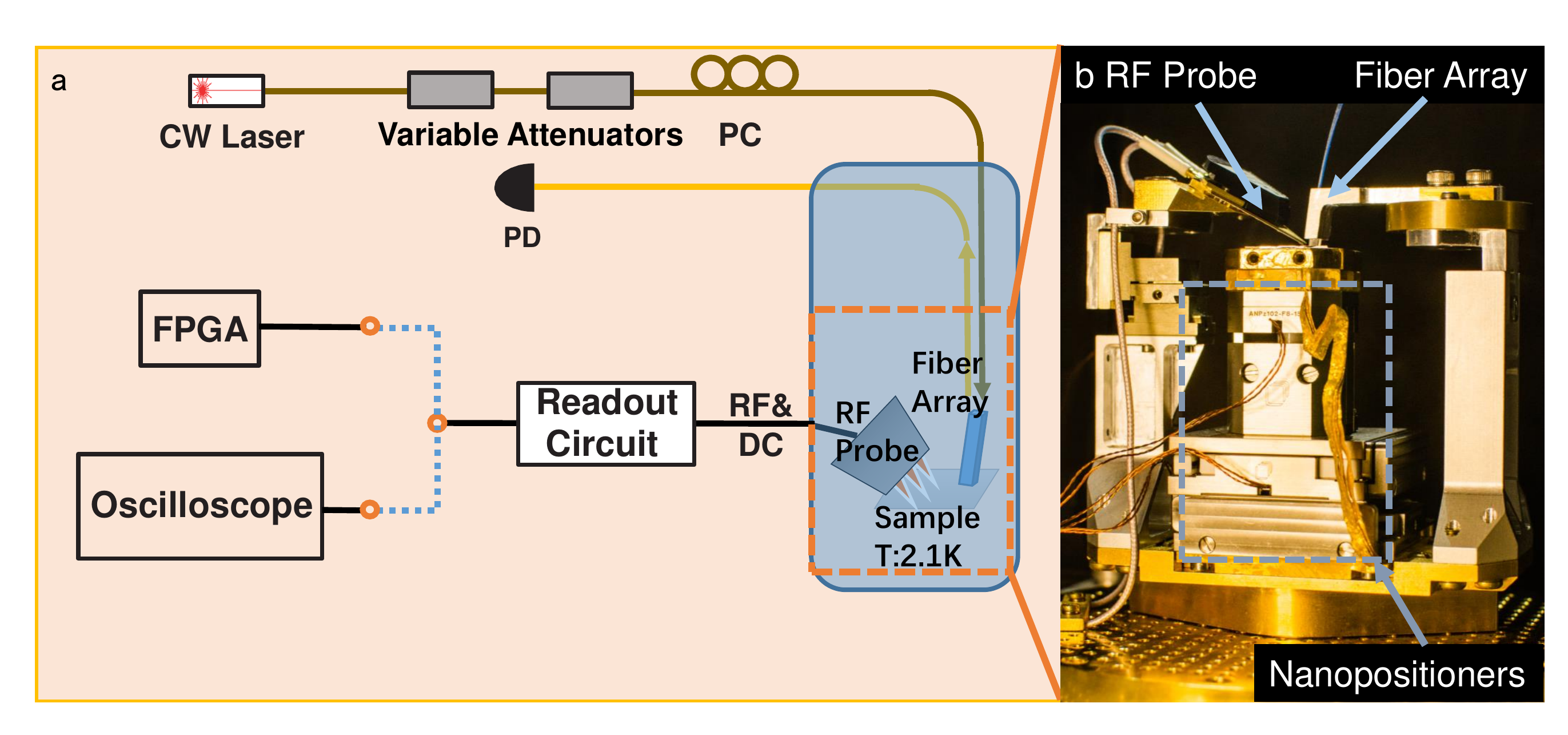}
    \caption{\label{Fig4} (a) Setup for critical current, OCDE and DCR measurements. (b) Photograph of sample stage which is mounted on a 3-axis closed-loop nano-positioner. See text for details.}
\end{center}
\end{figure*}

The fabrication steps of waveguide-integrated SNSPD are shown in Fig. \ref{Fig3} and explained in its caption. We use the setup in Fig. \ref{Fig4} to test the performance of the SNSPD, including its critical current, on-chip detection efficiency (OCDE) and dark count rate (DCR) in a closed-cycle fridge. We use a commercial electrical readout circuit to set bias current to detectors and obtain the response signal of detectors. Our sample is glued on the sample stage which is mounted on a 3-axis closed-loop low-temperature piezo nanopositioner. With the help of the nanopositioners, light can be coupled into the chip from fiber array. On the opposite side of fibre array, a RF probe is used to obtain the detectors' response signal.

After cooling the sample stage temperature down to 2.1 K, we perform further low-temperature measurements, such as critical current, temporal response, OCDE and DCR. In Fig. \ref{Fig5}, the I-V curves are measured by sweeping the voltage source between -1 V and 1 V. Each nanowire exhibits a superconducting state along the zero-voltage and abruptly transits to a normal-conducting state once the current is higher than critical current. It is clear to see that as the width of nanowire increases, the critical current increases.

\begin{figure*}[htbp]
\begin{center}
    \includegraphics[width=0.8\textwidth]{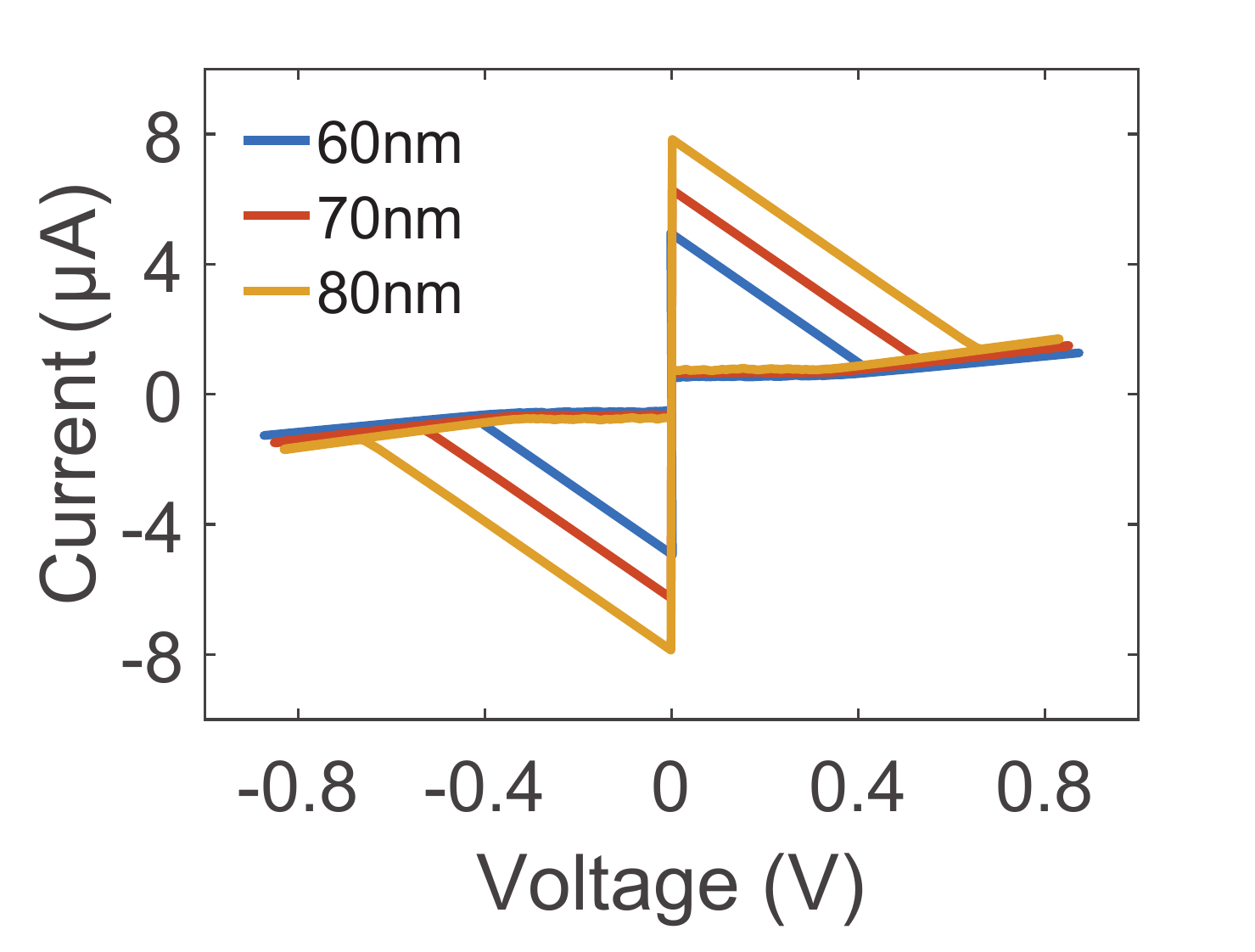}
    \caption{\label{Fig5} I-V curves of 3 nanowires with different widths. }
\end{center}
\end{figure*}

\begin{figure*}[htbp]
\begin{center}
    \includegraphics[width=0.8\textwidth]{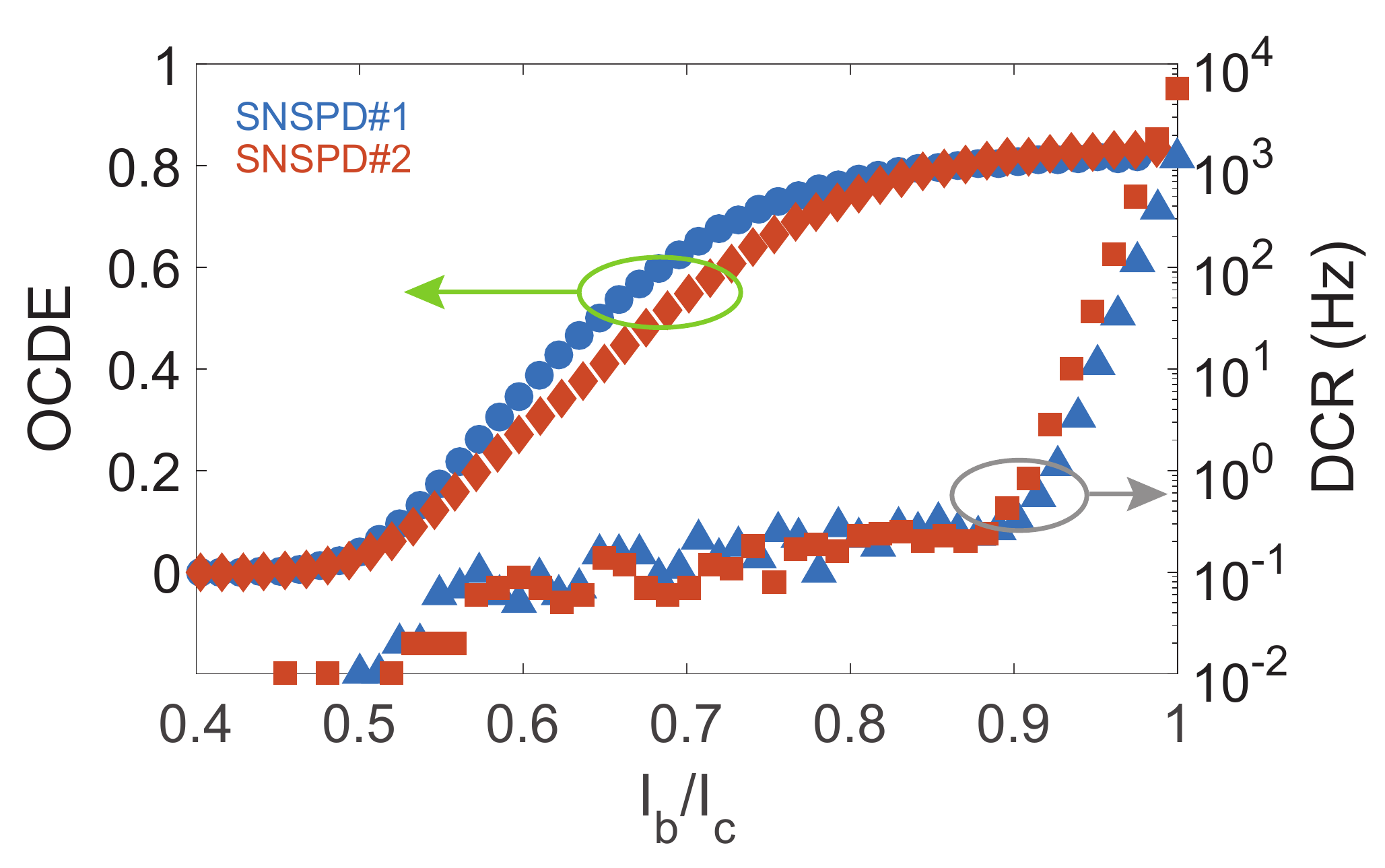}
    \caption{\label{Fig6} The OCDE and the DCR as a function of $I_b$/$I_c$ for the SNSPDs. $I_c$ is the critical current of SNSPDs. The data of OCDE and DCR are the averages of 100 one-second measurements.}
\end{center}
\end{figure*}

As shown in Fig. \ref{Fig4}, the OCDE of SNSPD is characterized by sending an attenuated continuous wave (CW) laser and the DCR is measured when laser is turn-off. A fiber polarization controller (PC) is used to maximize the input power. First, we characterize the coupling efficiency of the reference grating coupler. After the photon flux is guided to fiber array in the cryostat, we measure the input power $P_{in}$ and output power $P_{out}$ with the attenuator is set at 0 dB. The efficiency of the grating coupler can be calculated by the following formula
\begin{equation} \label{GrindEQ1}
\eta_{in}*\eta_{out}=\frac{P_{out}}{P_{in}},
\end{equation}
where $\eta_{in}$ and $\eta_{out}$ are the efficiency of the incidence grating coupler and emergent grating coupler, respectively. We neglect the on-chip propagation loss, so $\eta_{in}=\eta_{out}=\eta$. As derived from Fig. \ref{Fig2}, $\eta$ = -2.24 dB (1536nm). We also assume that the coupling efficiency of the main device's and the reference device's grating couplers are same. By adjusting the attenuator, we can find the photon number $N$ arriving at the detector in a similar way, with
\begin{equation} \label{GrindEQ2}
N=\frac{P_{in}}{h\nu}*\eta*r.
\end{equation}
where $r$ is the splitting ratio of MMI. We assume that the splitting ratio of multi-mode interference (MMI) is 50:50 in the following calculation. When the attenuation is $A$, with $N_{in}=N*A$, the on-chip detection efficiency (OCDE) can be written as
\begin{equation} \label{GrindEQ3}
OCDE=\frac{C-C_d}{N_{in}},
\end{equation}
where $C$ is OCCR (on-chip count rate) and $C_d$ is DCR. We can get the value of $C$ and $C_d$ by using a timetagger (UQD). DCR is the count per second when we turn off the laser.
We characterize the performance of our SNSPDs including OCDE and DCR with various bias currents at the wavelength of interests (1536nm). The results are shown in Fig. \ref{Fig6}. We change the bias current of our SNSPDs and obtain the OCDE and DCR as a function of normalized bias current. Saturation of count rate means that the internal quantum efficiency is close to 100 \%. In our QKD experiment, SNSPD\#1 is biased at 7.1 $\mu$A, where OCDE is about 0.80 and the dark count rate is about 0.25 Hz; SNSPD\#2 is biased at 6.8 $\mu$A, where OCDE is about 0.81 and the dark count rate is about 0.24 Hz.
\begin{figure*}[htbp]
\begin{center}
    \includegraphics[width=0.8\textwidth]{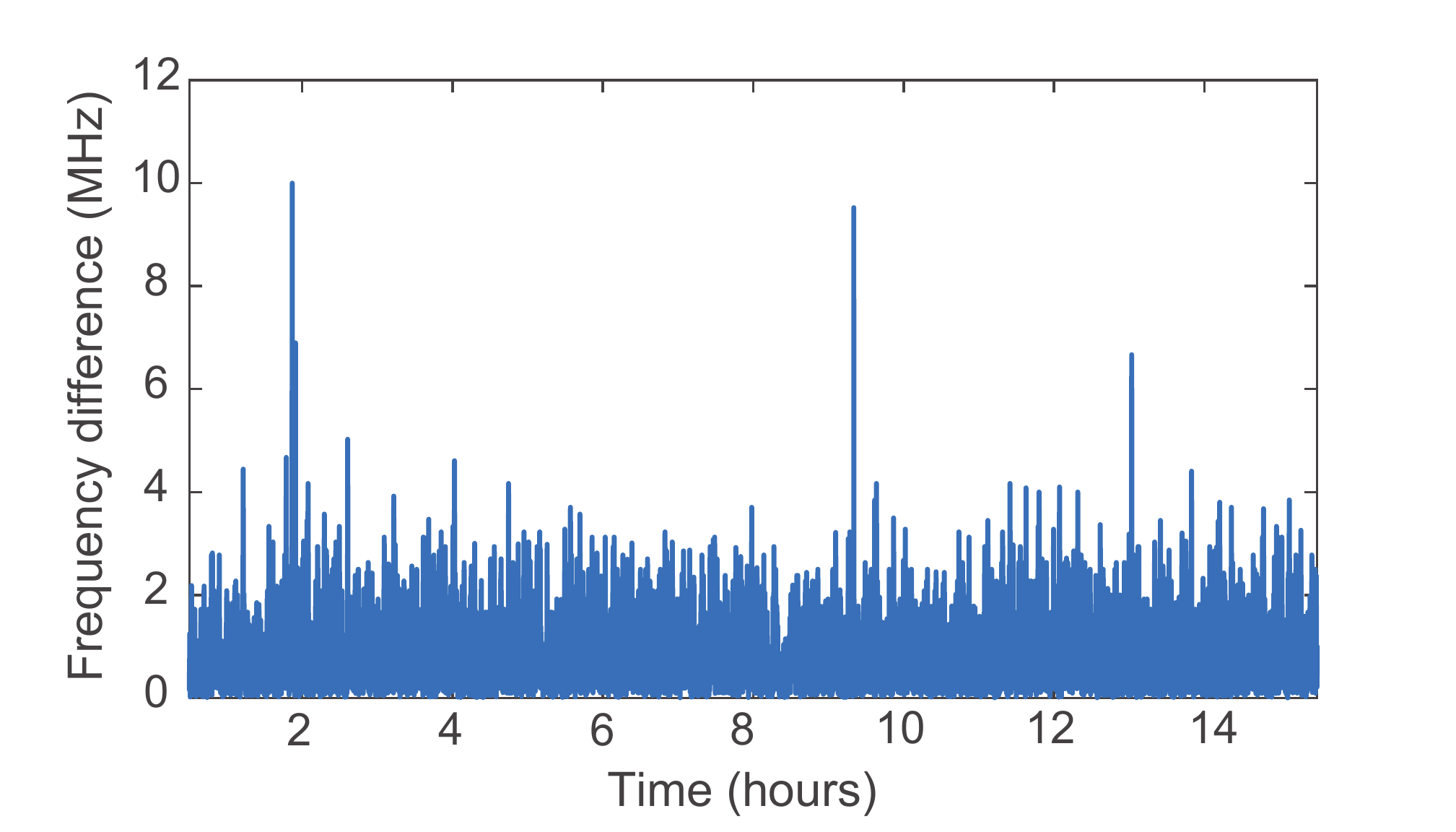}
    \caption{\label{Fig7} Frequency difference between the lasers of Alice and Bob with the feedback control. A maximum frequency drift no more than 10 MHz is observed in $\sim$15-hour monitoring data. }
\end{center}
\end{figure*}

\begin{figure*}[htbp]
\begin{center}
    \includegraphics[width=0.8\textwidth]{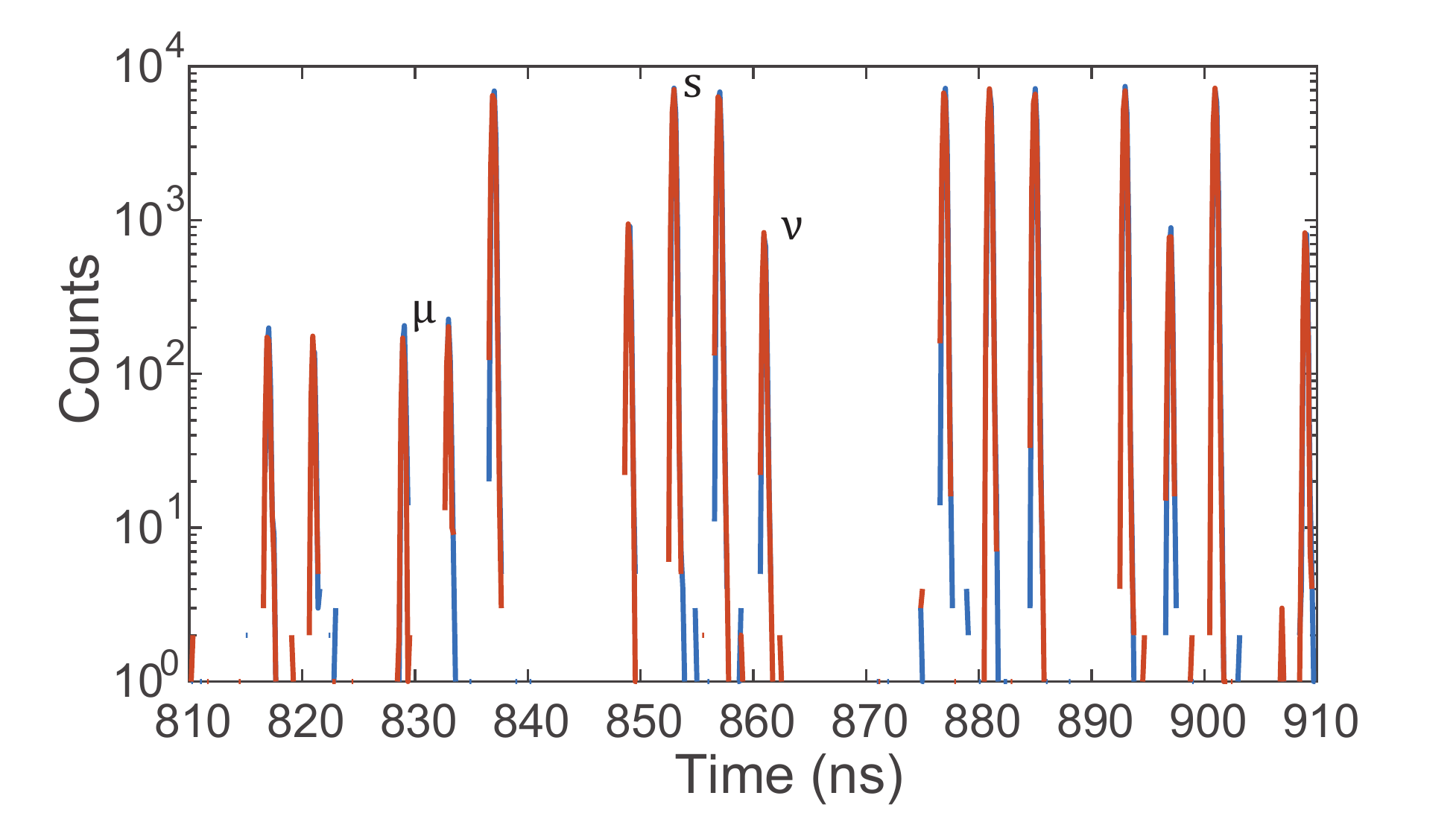}
    \caption{\label{Fig8} Histogram of the weak coherent pulse train (including signal state $s$, decoy state $\mu$ and $\nu$.) with a rate of 125 MHz is detected by the waveguide-integrated SNSPDs. The extinction ratio of the pulses is more than 20 dB. Red line and blue line are histogram between reference signal and two SNSPDs respectively. The detector electrical signals are collected by a timetagger with a temporal resolution of 156 ps. Count integration time is 200 s.}
\end{center}
\end{figure*}

\begin{figure*}
\begin{center}
    \includegraphics[width=0.8\textwidth]{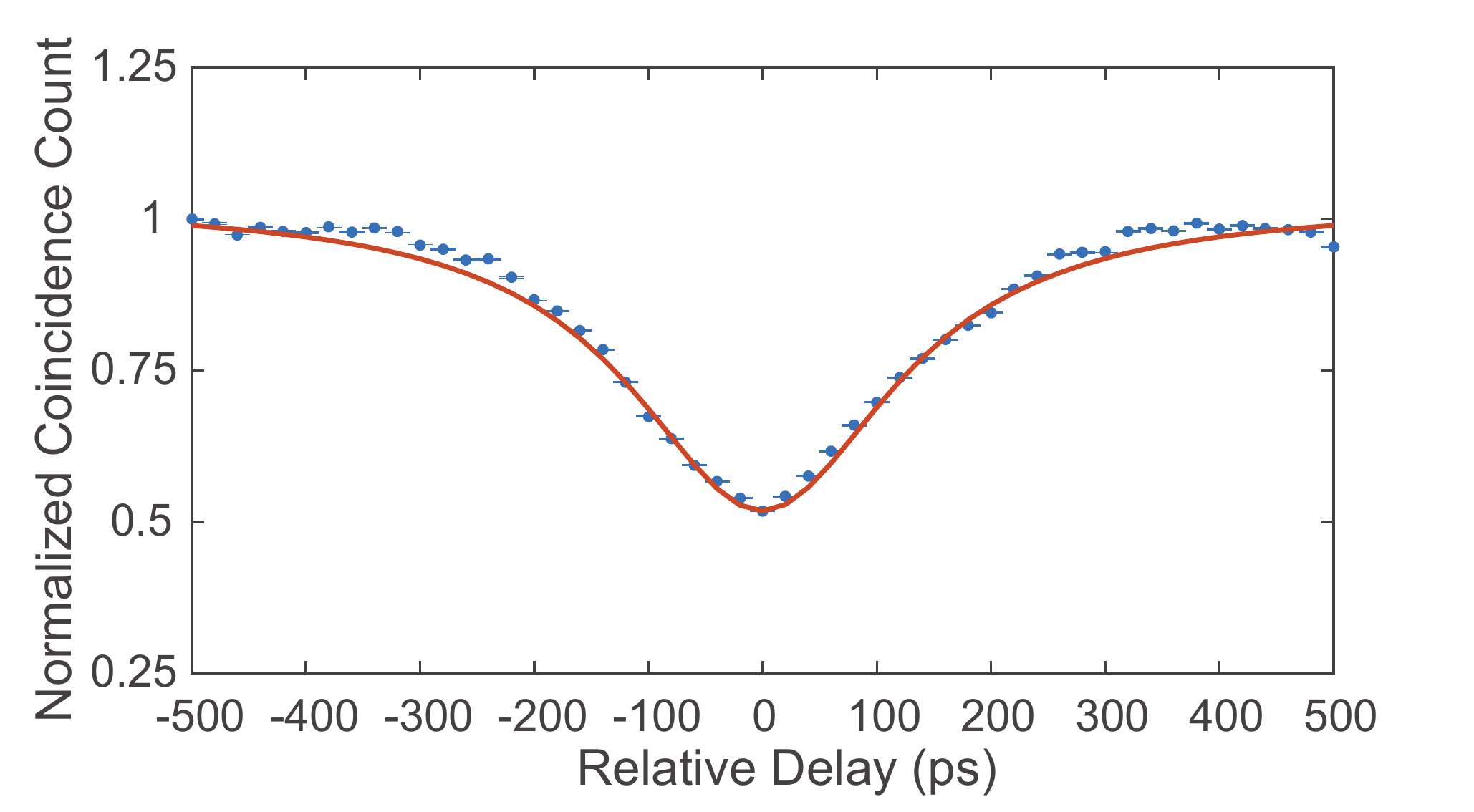}
    \caption{\label{Fig9}  Hong-Ou-Mandel (HOM) interference as a function of relative electronic delays between Alice and Bob. A HOM interference visibility of $V=48.1\%\pm 0.5\%$ is observed. The red curve is a fit. }
\end{center}
\end{figure*}
\subsection{Off-chip optical setup}\label{sec:two}
In our experiment, the light sources are two CW lasers with a nominal linewidth of 100 Hz centered at 1536.47 nm. For our experiment, it is important to keep the frequencies of both Alice's and Bob's lasers to be the same. We keep track on the frequency difference between these two lasers by beat measurement and employ a feedback system to regulate Bob's laser's frequency. The maximum frequency difference is less than 10 MHz over about 15 hours with the help of the feedback system, as shown in Fig. \ref{Fig7}. 

For each encoder module, we use one intensity modulator to chop out the required short pulses with about 370 ps full-width at half maximum (FWHM). Then we use another intensity modulators to modulate the mean photon number per pulse for decoy states ($\mu$ and $\nu$) and vacuum state ($o$). In Fig. \ref{Fig8}, we show the results of weak coherent pulses measured by our SNSPDs, in which a $>$20 dB extinction ratio has been obtained.

\subsection{Four intensity decoy-state analysis}\label{sec:three}
We use the four-intensity decoy-state protocol$^{\href{URL will be inserted by publisher}{46}}$ and consider the symmetric case
where Alice and Bob have equal channel transmissions to Charlie.
Experimentally, Alice and Bob each use one signal intensity $s$ in $Z$ basis for
key generation and three decoy intensities ($\mu$, $\nu$, $o$)
in $X$ basis for error test, where $o$ represents vacuum.
In practical implementation, the key size is finite and the statistical fluctuation should be considered
differently if pulse pairs from Alice ($l$) and Bob ($r$) in different
intensity combinations $lr$, with $l$, $r$$\in$\{$s$, $\mu$, $\nu$, $o$\}.
The key idea of the finite size analysis is to estimate a lower bound for the yield of single-photon pairs ($[y_{11}]_{min}$) and an
upper bound for the phase error rate of single-photon pairs ($[e_{11}^{ph}]_{max}$) from gain of the pulse pairs in the $X$ basis 
by applying the Chernoff bound to perform parameter estimation$^{\href{URL will be inserted by publisher}{47,48}}$. According to Ref.$\href{URL will be inserted by publisher}{46}$,
by jointly considering the observed data in $lr$$\in$$D$=\{$oo$, $o\mu$, $\mu o$, $o\nu$, $\nu o$, $\mu \mu$, $\nu \nu$\},
one can acquire tighter bound from the joint constrains than independent constrains, and hence produce
a higher key rate.
The final key rate can be extracted by 
\begin{equation}
R=p_sp_s\{s^2e^{-2s}[y_{11}]_{min}[1-h([e_{11}^{ph}]_{max})]-fy_{ss}h(E_{ss})\},
\end{equation}
where $p_sp_s$ is the probability of both Alice and Bob sending signal states.
$h(\cdot)$ is the Shannon binary entropy function, and $f=1.16$ is the efficiency of
error correction. $y_{ss}$ and $ E_{ss}$
denote the gain and QBER for $ss$ source, which can be directly obtained in experiments.

\subsection{Experimental results}\label{sec:four}

As shown in Fig. \ref{Fig9}, we obtain a HOM interference visibility of V=48.1\%$\pm$ 0.5\% with 4 ns time-bin separation at a rate of 250 MHz, indicating the high quality of our integrated chip for MDI-QKD. The blue dots are the experimental data and the red curve is a fit.

\begin{figure*}
\begin{center}
    \includegraphics[width=1\textwidth]{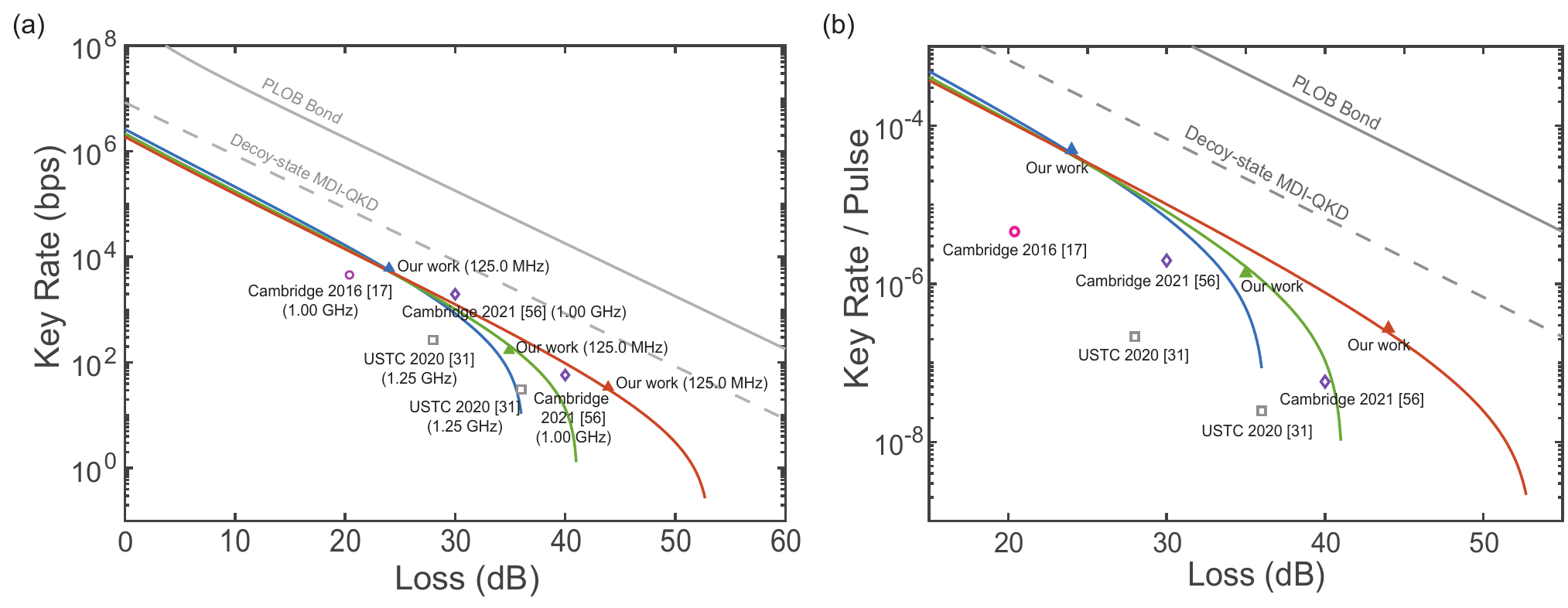}
    \caption{\label{Fig10} Our experiment results and comparison of state-of-the-art MDI-QKD experiments. (a) Key rate versus loss. (b) Key rate per pulse versus loss. }
\end{center}
\end{figure*}

 We employ our system to implement a complete MDI-QKD system including four intensity decoy states and phase randomization.In Table \ref{tab:1}, we show the mean photon number and proportion of different states and the sent total pulse pairs number at different loss.
 In Table \ref{tab:2}, we show detailed experimental results. The system has a 125 MHz repetition rate. 
 The system is run for 6.68 hours at total attenuations of 24.0 dB (including chip insertion loss $\sim$ 4.5 dB, the actual transmission loss corresponds to 98 km standard fibre),
 and a total of $3\times 10^{12}$ pulse pairs are sent from each sender. The key rate is 6166 bps. The system is also run for 6.68 hours at total attenuations of 35.0 dB (including chip insertion loss $\sim$ 4.5 dB, the actual transmission loss corresponds to 153 km standard fibre),
 and a same total of $3\times 10^{12}$ pulse pairs are sent from each sender. The key rate is 170 bps. 
 As for higher attenuation, 44.0 dB (including chip insertion loss $\sim$ 4.5 dB, the actual transmission loss corresponds to 198 km standard fibre), the run time is increased by one order (66.8 hours). So the sent pulse pairs number becomes $3\times 10^{13}$. The key rate is 34 bps. In Table \ref{tab:3}, We show the comparison of state-of-the-art MDI-QKD experiments. We emphasize that our secure key rates with 125~MHz clocked system are very close to the best MDI-QKD experiments with GHz clock rate$^{\href{URL will be inserted by publisher}{17,31,56}}$.
In Fig. \ref{Fig10}, we set the clock rate equal to 125 MHz for all the simulated rates. To draw these curves, we have set transmissivity $\eta = 10^{-Loss(dB)/10}$. The PLOB curve and the decoy-state MDI-QKD curve$^{\href{URL will be inserted by publisher}{57}}$ have been obtained under ideal conditions, i.e., with zero detector and channel noise, maximum detection efficiency (detection efficiency is 1) and unitary error correction efficiency ($f = 1$). 
PLOB bond: $R_P = -log_2(1-\eta)$.
Decoy-state MDI-QKD: $R_D = \eta/(2e^2)$.
\begin{table}[!htbp]
\centering
\caption{List of mean photon number and proportion of signal state and decoy state}
\label{tab:1}
\begin{threeparttable}
\begin{tabular}{| l | c | c | c | c | }
	\hline
Loss\tnote{*} (dB) & 24.0 & 35.0 & 44.0 \\ \hline
$s$ & 0.714 & 0.66 & 0.624 \\ \hline
$\mu$ & 0.034 & 0.048 & 0.054  \\ \hline
$\nu$ & 0.172 & 0.196 & 0.208 \\ \hline
$P_s$ & 0.828 & 0.774 & 0.736 \\ \hline
$P_{\mu}$ & 0.14 & 0.176 & 0.204 \\ \hline
 $P_{\nu}$ & 0.014 & 0.03 & 0.039 \\ \hline
$N$ & $3\times 10^{12}$ & $3\times 10^{12}$ & $3\times 10^{13}$ \\ \hline
	
\end{tabular}
\begin{tablenotes}
\footnotesize
\item[*] Including chip insertion loss $\sim$ 4.5 dB
\end{tablenotes}
\end{threeparttable}
\end{table}

\begin{table}[!htbp]
\centering
\caption{List of the total gains and error gains of Bell states. The notion $N_{ij}$
denotes the number of pulse pairs sent out from $ij$ intensity combination.}
\label{tab:2}
\begin{threeparttable}
\begin{tabular}{| l | c | c | c | c | }
	\hline
Loss\tnote{*} (dB) & 24.0 & 35.0 & 44.0 \\ \hline
$N_{ss}y_{ss}$ & 1340443872 & 81820241 & 83430549 \\ \hline
$N_{ss}y_{ss}E_{ss}$ & 89872 & 8187 & 29458  \\ \hline
$N_{\mu\mu}y_{\mu\mu}$ & 189673 & 47419 & 87324 \\ \hline
$N_{\mu\mu}y_{\mu\mu}E_{\mu\mu}$ & 54435 & 13193 & 24656 \\ \hline
$N_{\nu\nu}y_{\nu\nu}$ & 33781 & 25250 & 37087 \\ \hline
 $N_{\mu o}y_{\mu o}+N_{o \mu}y_{o \mu}$ & 11123 & 2673 & 4632 \\ \hline
  $N_{\nu o}y_{\nu o}+N_{o \nu}y_{o \nu}$ & 17115 & 6481 & 9494 \\ \hline
  $N_{oo}y_{oo}$ & 0 &  0 & 0 \\ \hline
  $y_{11}$ & $1.60\times 10^{-3}$ & $9.29\times 10^{-5}$ & $1.28\times 10^{-5}$\\ \hline
  $e^{ph}_{11}$ & 0.1616 & 0.2321 & 0.1557\\ \hline
  Key rate/ pulse & $4.93\times 10^{-5}$ & $1.36\times 10^{-6}$ & $2.72\times 10^{-7}$
\\
\hline
\end{tabular}
\begin{tablenotes}
\footnotesize
\item[*] Including chip insertion loss $\sim$ 4.5 dB
\end{tablenotes}
\end{threeparttable}
\end{table}

\begin{table}[!htbp]
\centering
\caption{Comparison of state-of-the-art MDI-QKD experiments.}
\label{tab:3}
\begin{threeparttable}
\begin{tabular}{ l  c  c  c  c  }
\toprule
Reference & Clock rate (MHz) & Loss (dB) & Key rate (bps) & Key rate/ pulse\\
\midrule
Comandar \textit{et al.}$^{\href{URL will be inserted by publisher}{17}}$ & 1000 & 20.4 & 4567\tnote{a} & $4.57\times 10^{-6}$ \\ \hline
Wei \textit{et al.}$^{\href{URL will be inserted by publisher}{31}}$ & 1250 & 20.4 & 6172\tnote{b} & $4.94\times 10^{-6}$\\
 &  & 28.0 & 268 & $2.14\times 10^{-7}$\\
 &  & 36.0 & 31 & $2.48\times 10^{-8}$\\ \hline
Woodward \textit{et al.}$^{\href{URL will be inserted by publisher}{56}}$ & 1000 & 30.0 & 1971 & $1.97\times 10^{-6}$\\ 
 &  & 40.0 & 58 & $5.80\times 10^{-8}$\\ \hline
This work  & 125 & 24.0\tnote{c} & 6166 & $4.93\times 10^{-5}$\\ 
 &  & 35.0\tnote{c} & 170 & $1.36\times 10^{-6}$\\
 &  & 44.0\tnote{c} & 34 & $2.72\times 10^{-7}$\\
\bottomrule
\end{tabular}
\begin{tablenotes}
\footnotesize
\item[a] No random modulations.
\item[b] Simulation.
\item[c] Including chip insertion loss $\sim$ 4.5 dB.
\end{tablenotes}
\end{threeparttable}
\end{table}

\clearpage
\end{document}